\documentclass[a4paper]{article}
\usepackage{graphicx}
\usepackage{amsmath}
\usepackage{amssymb}

\newcommand{\EUV}{{\epsilon}}

\newcommand{\Set}[1]{\{#1\}}

\newcommand{\SetIn}[2]{\{#1 \;|\; #2 \}}

\newcommand{\SetInteger}{\mathbb{Z}}

\newcommand{\SetReal}{\mathbb{R}}

\newcommand{\Eq}{\;=\;}

\newcommand{\DualCone}[1]{{#1}^{\vee}}
\newcommand{\DualDualCone}[1]{{#1}^{\vee\vee}}

\newcommand{\Norm}[1]{\parallel{#1}\parallel}

\newcommand{\Polyn}[1]{{\cal #1}}

\newcommand{\LCM}{\mathrm{lcm}}

\newcommand{\ConvexHull}{\mathrm{conv}}

\newcommand{\Face}{\mathcal{F}}

\newcommand{\VOne}{\mathbb{I}}

\newcommand{\Comment}[1]{}

\begin{document}
\title{A geometric method of sector decomposition}
\author{Toshiaki {\sc Kaneko} \\
  \textit{High Energy Accelerator Research Organization (KEK),}\\
  \textit{Ohe 1-1, Tsukuba, Ibaraki 305-0801, Japan}
  \and Takahiro {\sc Ueda} \\
  \textit{Graduate School of Pure and Applied Sciences,} \\
  \textit{University of Tsukuba,}\\
  \textit{Tsukuba, Ibaraki 305-8571, Japan}
}
\date{}
\maketitle

\begin{abstract}
We propose a new geometric method of IR factorization in
sector decomposition.
The problem is converted into a set of problems in convex geometry.
The latter problems are solved using algorithms in
combinatorial geometry.
This method provides a deterministic algorithm and never falls into an
infinite loop.
The number of resulting sectors depends on the
algorithm of triangulation.
Our test implementation shows smaller number of sectors
comparing with other existing methods with iterations.
\end{abstract}

\section{Introduction}

Analysis of recent high statistical experiments in high energy physics
requires accurate theoretical predictions.
Higher order corrections in perturbation theory become
indispensable for the calculation of many physical processes.
However, calculation of higher loop integration is not an easy
problem because of its high dimensionality and complicated dependence on
many kinematic and mass parameters.
As the required final results of calculations are sets
of numerical values, various numerical
methods of loop integrations have been proposed.
Since numerical method can only be applied to the quantities without
divergences, one has to avoid singularities in the integrand by
regularization or subtraction of divergent part before
applying numerical method.

In the analysis of loop corrections in perturbative QCD,
\(D\)-dimensional regularization of infrared (IR) divergences
is commonly used, where
these divergences are separated in the form of
poles in terms of \(\EUV = (4-D)/2\).
The method of sector decomposition~%
\cite{phys:Hepp-1966,phys:Roth-Denner-1996,phys:Binoth-Heinrich-2000,%
      phys:Binoth-Heinrich-2004,phys:Binoth-Heinrich-2004a,phys:Heinrich-2008}
provides a systematic procedure to extract these poles,
and their coefficients are free from IR divergences.
This method consists of several steps, whose details will be discussed
in the next section. Here we mention each of these steps briefly:

\begin{enumerate}
\item Start from Feynman parameter representation of a loop integration.

\item Primary sector decomposition:

      Divide the domain of integration
      into sectors,
      eliminating the $\delta$-distribution,
      such that each sector includes one corner of 
      the original integration domain.

\item Factorization step:

      Decompose sectors repeatedly 
      such that IR diverging part
      of the integrand is factored out as 
      a product of powers of the integration variables
      in each sector.

\item Divergence separation step:

      When the integrand in each sector has a divergent factor
      of an integration variable, one can extract the singular part
      expanding into Laurent series.
      Poles of $\EUV$ are separated with the analytic integration of the
      singular part.
      After all poles are extracted by repeating this procedure for
      all divergent factors, 
      the finite part is expanded into Taylor series in terms of
      $\EUV$.

\item Integration step:

      By the previous step,
      the coefficients of poles and the finite part
      are expressed by integrals of
      functions without IR divergences.
      If these functions do not suffer from other singularities,
      these integrations are calculated numerically.
      Otherwise further regulations of other
      singularities are 
      necessary~%
      \cite{phys:Lazopoulos:2007ix,phys:Anastasiou-2007qb,phys:Pilipp-2008,phys:Ueda-Fujimoto-2008}

\end{enumerate}

The factorization step of the usual sector decomposition approach
decomposes the integration domain into sectors 
according to a \textit{sector decomposition strategy}.
A strategy provides a rule 
to choose a subset of integration variables
and to determine how chosen variables are transformed 
and how the domain is divided.
After applying these decompositions iteratively, 
one will obtain integrands whose IR divergent part
is correctly factored out in each resulting sector. 
However, it is found that some strategies may lead
to an infinite loop of iteration.
This problem was first solved
by Ref.~\cite{phys:Bogner-Winzier-2008,phys:Bogner-Winzier-2008a}
with presenting three strategies guaranteed to terminate.
Their strategies are based
on the analysis of the polynomials appearing in the integrand.
The strategy given in 
Ref.~\cite{phys:Smirnov-Tentyukov-2008,phys:Smirnov-Smirnov-2008}.
also solve this problem, which is
based on the analysis of polytopes corresponding to the polynomials.

The number of generated sectors depends heavily
on the strategies.
The efficiency of sector decomposition method depends
on the resulting number of
decomposed sectors, since the numerical integration is the most
time consuming part and will
be repeated for different values of kinematic or mass parameters.
Therefore such a method of the factorization step is favored
that gives as small a number of generated sectors as possible.

In this paper, we propose a new method of the factorization step
employing geometric interpretation of the problem.
This is a deterministic method without iterated decomposition and
hence never falls into an infinite loop.
In section~2, the problem is rephrased in clarifying the condition of
the expected output of the factorization step.
The problem is converted to a set of problems in convex geometry in section~3.
Latter problems are solved in section~4 with algorithms developed
in combinatorial geometry.
Then it is shown that the number of the decomposed sectors depends on the
triangulation algorithm of \(n\)-dimensional polytopes.
In section 5, we show the results of the test implementation
of our method.
The last section is devoted to the conclusion.

Throughout this article, we use the following `multi-index notation'.
For \(m\)-tuple of variables \(x = (x_1, \cdots, x_m)\) and numbers
\(a = (a_1, \cdots, a_m)\),
the product of \(x_j^{a_j}\) for \(j = 1, \cdots, m\) is expressed by
\begin{eqnarray}
x^{a} &:=& x_1^{a_1} \cdots x_m^{a_m} ,
\end{eqnarray}
and
\begin{eqnarray}
d^{m} x &:=& d x_1 \, \cdots \, d x_m .
\end{eqnarray}
We also use special vector \(\VOne\)
defined by \(\VOne := (1, \cdots, 1)\).

\section{Sector decomposition}

Let us start with Eq.~(11) of Ref.~\cite{phys:Heinrich-2008}
for the loop integration expressed with Feynman parameters
\(x = (x_1, \cdots, x_N)\):
\begin{equation}
\label{sect:0}
G = (-1)^{N_\nu} \frac{\Gamma(N_\nu - L D/2)}{
                         \prod_{j=1}^N \Gamma(\nu_j)}
      \int_0^\infty d^N x \; x^{\nu-\VOne}
        \delta\left(1-\sum_{l=1}^N x_l\right)
        \frac{\Polyn{U}^{N_\nu - (L+1) D/2}}{
              \Polyn{F}^{N_\nu - L D/2}},
\end{equation}
where
\begin{itemize}
\item \(L\) is the number of loops,
\item \(D = 4 - 2 \EUV\) is the dimension of the space-time,
\item \(N\) is the dimension of the integration,
\item \(\nu_j \; (j= 1,\cdots, N)\)
       is the power of the propagator
       corresponding to the Feynman parameter \(x_j\),
\item \(N_\nu := \sum_{j=1}^N \nu_j\),
\item \(\Polyn{U}\) is a homogeneous polynomial of
      \(\Set{x_j}\) of degree \(L\), and
      all the coefficients of the monomials of \(\Polyn{U}\) are equal to 1.
\item \(\Polyn{F}\) is a homogeneous polynomial of
      \(\Set{x_j}\) of degree \(L+1\),
      and the coefficients of the monomials of $\Polyn{F}$ consist of
      kinematic and mass parameters.
\end{itemize}

Polynomial $\Polyn{F}$ may approach to 0 when the vector \(x\)
of Feynman parameters approaches to a corner of the integration domain
constrained by the $\delta$-distribution ($x_j \rightarrow 0$ except for
one parameter).
In such cases, the integration may diverge and cause IR singularities,
which are dimensionally regulated.

The primary sector decomposition step
is to separate these singularities with dividing the integration domain
into \(N\) sub-domains
such that each sector includes only one corner.
Sector \(l\) is defined by:
\begin{eqnarray}
\SetIn{(x_1,x_2,\cdots,x_N)}{x_j \leq x_l, \; \forall j \neq l} .
\end{eqnarray}
In this sector,
the integration variables are changed
from \(\Set{x_1, \cdots, x_N}\) to \(\Set{x_l, t_1, \cdots, t_{N-1}}\)
in the following way:
\begin{eqnarray}
x_j &=& x_l t'_j ,
\\
t'_j &:=& \left\{ \begin{array}{lll}
          t_{j}    & \mbox{for} & j < l \\
            1      & \mbox{for} & j = l \\
          t_{j-1}  & \mbox{for} & j > l
       \end{array} \right. .
\end{eqnarray}
Integrating over \(x_l\) in Eq.~(\ref{sect:0}),
one obtain Eq.~(14) of Ref.~\cite{phys:Heinrich-2008}:
\begin{eqnarray}
G &=& (-1)^{N_\nu} \frac{\Gamma(N_\nu - L D/2)}{
                         \prod_{j=1}^N \Gamma(\nu_j)}
      \sum_{l=1}^N G_l ,
\\
G_l &=&
      \int_0^1 d^{N-1} t \; t^{\nu'-\VOne}
        \;
        \Polyn{U}_l^{\gamma}(t) \, \Polyn{F}_l^{\beta}(t) ,
\label{sd:gl}
\end{eqnarray}
where
\begin{eqnarray}
\Polyn{F}_l(t) &=& \Polyn{F}(t'),
\qquad
  \beta \Eq - (N_\nu - L D/2) ,
\\
\Polyn{U}_l(t) &=& \Polyn{U}(t'),
\qquad
  \gamma \Eq N_\nu - (L+1)D/2 ,
\end{eqnarray}
and
\begin{equation}
  \nu'_j
  := \left\{
      \begin{array}{lll}
        \nu_j & \mbox{for} & j < l \\
        \nu_{j+1} & \mbox{for} & j \ge l
      \end{array}
    \right. .
\end{equation}
After the primary sector decomposition,
the domain of integration becomes $(N-1)$-dimensional unit cube and
the corner of the original domain of integration is mapped to
the origin
$t_j = 0$ (\(\forall j\)), where $\Polyn{U}_l$ and $\Polyn{F}_l$ 
may vanish.
This expression is the starting point of our discussion.

The purpose of the factorization step is
factoring out the possible singularities at $t_j = 0$ from
\(\Polyn{U}_l\) and \(\Polyn{F}_l\).
In the usual sector decomposition approach,
this factorization is obtained by dividing the domain of integration
repeatedly and applying appropriate transformation of integration variables.
With an appropriate strategy, one can finally find proper sub-domains
and new variables suitable for the separation of IR divergences.
To be more specific:
\begin{enumerate}
\item Integration domain is divided into \(m\) sectors
      \(\Set{D_1, D_2, \cdots, D_m}\).
\item In each sector \(D_a\),
      new variables \(z = (z_1,\cdots, z_{N-1})\) are introduced
      and \(t_j = t_j(z)\) is expressed as a monomial of \(z\), and
      Jacobian \(J_a(z)\) is a monomial of \(z\)
      with constant coefficient.
      Let monomial \(z^{c_a}\) be
      a product of $t^{\nu'-\VOne}$ and Jacobian:
\begin{eqnarray}
      z^{c_a} \Eq \prod_j z_j^{(c_a)_j} &:=& t(z)^{\nu'-\VOne} \, J_a(z) 
      \Eq \prod_k t_k(z)^{\nu_{k'}-1} J_a(z) .
\end{eqnarray}
\item Integration domain of \(z\) is the (\(N-1\))-dimensional
      unit cube (\(0 \leq z_j \leq 1\)).
\item Polynomials \(\Polyn{U}_l\) and \(\Polyn{F}_l\)
      are expressed in the following form in each sector \(D_a\):
\begin{eqnarray}
    \Polyn{U}_l &=&
    C_a z^{b_{a}}
              \left( 1 + H_a(z)\right),
\\
    \Polyn{F}_l &=&
    C_a' z^{b_{a}'}
              \left( 1 + H_{a}'(z)\right),
\label{sd:fact}
\end{eqnarray}
     where \(b_{a}\) and \(b_{a}'\)
     are \((N-1)\)-tuples of non-negative integers
     (\(b_a, b_a' \in \SetInteger_{\geq 0}^{N-1}\))%
\footnote{%
In this paper,
the set of non-negative integers and the set of non-negative
real numbers are denoted by
\(\SetInteger_{\geq 0}, \SetReal_{\geq 0}\),
respectively.
}%
,
     \(H_a(z)\) and \(H_a'(z)\)
     are polynomials of \(z\) such that
     \(H_a(0) = 0\) and \(H_a'(0) = 0\).
\end{enumerate}
The output of the factorization step is shown by the
following form of the expression:
\begin{eqnarray}
G_l &=&
        \sum_a C_a'^\beta C_a^{\gamma}
      \int_0^1 d^{N-1} z \; z^{c_a + b_{a}' \beta + b_{a}\gamma}
        \;
\\ \nonumber & & \qquad\qquad\times
              \left( 1 + H_a(z)\right)^\gamma
              \left( 1 + H_a'(z)\right)^\beta .
\end{eqnarray}

The divergence separation step proceeds in the following way.
Factor \(z^{c_a + b_{a}' \beta + b_{a}\gamma}\) is expressed by
\(\prod_j z_j^{n_j + m_j \EUV} \; (n_j, m_j \in \SetInteger)\).
If \(n_j \leq -1\),
the integration of \(z_{j}\) diverges
when $\EUV\to0$.
In this case, the integrand is split into IR divergent and finite
parts.
Let us consider the case of \(n_j=-1\) for simplicity:
\begin{eqnarray}
& &
      \int_0^1 d z_j \; z_j^{ -1 + m_j \EUV} \;
              \left( 1 + H_a(z)\right)^\gamma
              \left( 1 + H_a'(z)\right)^\beta
\nonumber
\\ & & \qquad \Eq
      \int_0^1 d z_j \; z_j^{ -1 + m_j \EUV} \;
              \left( 1 + H_a(z)|_{z_j=0}\right)^\gamma
              \left( 1 + H_{a'}(z)|_{z_j=0}\right)^\beta
\\ \nonumber & & \qquad \quad
    + \int_0^1 d z_j \; z_j^{ -1 + m_j \EUV} \;
       \Big[
              \left( 1 + H_a(z)\right)^\gamma
              \left( 1 + H_{a'}(z)\right)^\beta
\\ \nonumber & & \qquad \qquad \qquad \qquad
          -
              \left( 1 + H_a(z)|_{z_j=0}\right)^\gamma
              \left( 1 + H_{a'}(z)|_{z_j=0}\right)^\beta
      \Big] .
\end{eqnarray}
The first term is integrated to
\begin{eqnarray}
& &
   \frac{1}{ m_j \EUV}
              \left( 1 + H_a(z)|_{z_j=0}\right)^\gamma
              \left( 1 + H_{a'}(z)|_{z_j=0}\right)^\beta ,
\end{eqnarray}
and the
integrand of the second term is finite for \(z_j \rightarrow 0\).
When \(n_j < -1\), it is necessary to expand
\(
    \left( 1 + H_{a}(z)\right)^\gamma
    \left( 1 + H_{a'}(z)\right)^\beta
\)
around \(z_j = 0\) up to the necessary order
for the separation of the IR divergent parts.
Repeating this procedure for all divergent integrals, we can extract all
IR divergent parts as poles in terms of $\EUV$. 
Then we can freely expand integrands of remaining integrals with respect
to $\epsilon$, and obtain a Laurent series of the whole integral in
terms of $\epsilon$, whose coefficients are expressed by IR
finite integrals.
We can calculate these IR finite integrals numerically if
there are no other singularities in the integrands.

Let us go back to the factorization step.
In order to see the problem in a different way,
let us apply the expression~(\ref{sd:fact}) to a simple polynomial case
of three variables \(t_1, t_2, t_3\):
\begin{eqnarray}
\Polyn{F}_l(t) = - s_{12} t_1 - s_{23} t_2 t_3 - s_4 t_1 t_3 .
\label{sd:example}
\end{eqnarray}
When the integration variables are changed to \(z_1, z_2, z_3\) by
\begin{eqnarray}
t_1 = z_1 z_2 z_3, \qquad t_2 = z_1, \qquad t_3 = z_2,
\end{eqnarray}
we obtain
\begin{eqnarray}
\Polyn{F}_l(t(z)) = z_1 z_2 ( - s_{23} - s_4 z_2 z_3 - s_{12} z_3) .
\end{eqnarray}
This expression implies that around the origin of the domain
\(\SetIn{(t_1(z), t_2(z), t_3(z))}{0 \leq z_i \leq 1}\),
the term \(s_{23} t_2 t_3 = s_{23} z_1 z_2 \) dominates to the others
and the ratios of other terms to the dominant one
become monomials of the new variables \(z_i\).
Similarly,
changing variables in Eq.~(\ref{sd:example}) as
\begin{equation}
  t_1 = z_1 , \qquad t_2 = z_1 z_2 , \qquad t_3 = z_3 ,
\end{equation}
gives us
\begin{equation}
  \Polyn{F}_l(t(z)) = z_1 (- s_{12} - s_{23} z_2 z_3 - s_4 z_3) ,
\end{equation}
and this exhibits a case that $s_{12} t_1 = s_{12} z_1$ is
dominant at the origin.
Note that $s_4 t_1 t_3$ cannot be a dominant term at the origin because
always $|s_{12} t_1| > |s_4 t_1 t_3|$ when all $t_j$ approach to 0.

Above example shows that the factorization step
is equivalent to solve the following set of problems:
\begin{enumerate}
\item to determine which terms can be dominant,
\item to determine the sub-domain where each term becomes dominant, and
\item to find a new parameterization of the variables such that
      other terms are expressed as monomials multiplied by the
      dominant term.
\end{enumerate}

In the following sections, we will show that these problems can be solved
by a new method without iterations.

\section{Geometric method}

Let us change the integration variables from
\(t = \Set{t_1, \cdots, t_{N-1}}\) to \(y = \Set{y_1, \cdots, y_{N-1}}\)
in Eq.~(\ref{sd:gl}) as the following:
\begin{equation}
  t_j = e^{-y_j} \qquad \mbox{or} \qquad y_j = - \log t_j \qquad
  (j = 1, \cdots, N-1).
\end{equation}
Jacobian is \(e^{- \sum_j y_j}\) and
the integration domain is changed from
\(0 < t_j < 1\) to \(0 < y_j < \infty\).
We obtain
\begin{eqnarray}
G_l &=& \int_0^\infty d^{N-1} y \;
         e^{- (\nu', y)}  \,
         \Polyn{U}_l^{\gamma}(e^{-y}) \,
         \Polyn{F}_l^{\beta}(e^{-y}) ,
\end{eqnarray}
where
\((\nu', y)\) is the inner product defined in
\((N-1)\)-dimensional Euclidean space.

A monomial $t^b = t_1^{b_1} t_2^{b_2} \cdots t_{N-1}^{b_{N-1}}$
is characterized by an integer vector
\(b = \Set{b_1, \cdots, b_{N-1}}\; (b_j \in \SetInteger_{\geq 0}^{N-1})\).
Since a polynomial \(P\) is a sum of monomials,
the polynomial
corresponds to
a finite set of such vectors 
\(Z^P \subset \SetInteger_{\geq 0}^{N-1}\).
In terms of the new variables \(\Set{y_j}\),
\(P\) is expressed by:
\begin{eqnarray}
P(t) = \sum_{b \in Z^P} a_b t^b = \sum_{b \in Z^P} a_b e^{-(b, y)}
\qquad (a_b \neq 0,  \; \forall b \in Z^P) .
\label{geo:pt}
\end{eqnarray}

Let us consider the limit \(\lambda \rightarrow + \infty\) for
the variables \(y_j = \lambda u_j \rightarrow + \infty \; (\forall j)\)
with a constant vector \(u \in \SetReal_{\geq 0}^{N-1} \backslash \{0\}\).%
\footnote{\(A \backslash B = \SetIn{a \in A}{a \not\in B}\).
Do not confuse with Minkowsky sum \(A - B = A + (-B)\) defined
by Eq.~(\ref{minkowsky})}
The dominant term of \(P(t)\) 
in this limit
is the monomial \(a_b t^b\)
whose value of \((b, u)\) takes the minimum value of
\((c,u)\) among all \(c \in Z^P\):
\begin{eqnarray}
(b, u) &=& \min \SetIn{(c, u)}{c \in Z^P}.
\label{geo:condv}
\end{eqnarray}
In other words, a monomial for a vector \(b\) is dominant
in the domain \(\Delta_b^P\) of \(y\) space, which is defined by:
\begin{eqnarray}
\Delta_b^P &:=&
  \SetIn{y \in \SetReal_{\geq 0}^{N-1}}{(b, y) \leq (c, y) , \; \forall c \in Z^P} .
\label{geo:dom}
\end{eqnarray}
Evidently the union of all \(\Delta_b^P\) covers whole integration
domain.
The polynomial \(P\) of Eq.~(\ref{geo:pt}) is rewritten to:
\begin{eqnarray}
P(t(y)) &=&
       \sum_{b \in Z^P}
       \theta(y \in \Delta_b^P) \;
         e^{-(b, y)} \left[
          a_b + \sum_{c \in Z^P \backslash \Set{b}} a_c e^{-(c-b, y)}
         \right] .
\end{eqnarray}
Since \((c-b, y) \geq 0\) for \(y \in \Delta_b^P\),
the behavior of \(P\) in the limit of
\(\Norm{y} \rightarrow \infty\)
is determined by the term \(a_b e^{-(b, y)}\).
If a monomial corresponding to $b$ cannot be a dominant term, 
$\Delta_b^P$ is the empty set.
Including this case,
when the dimension of \(\Delta_b^P\) is less than \(N-1\),
the (\(N-1\))-dimensional volume of \(\Delta_b^P\) is zero.
We can ignore such a sub-domain, since it does not contribute to the
integration.
Applying the above discussion to $\Polyn{U}_l$ and $\Polyn{F}_l$ for $P$,
we obtain:
\begin{equation}
\label{geo:gly}
\begin{split}
G_l =&
      \sum_{b \in Z^{\Polyn{U}_l}}
      \sum_{b' \in Z^{\Polyn{F}_l}}
      \int_0^\infty d^{N-1} y \;
       \theta(y \in \Delta_{b b'}) \;
         e^{-(\nu' + \gamma b + \beta b', y)}  \,
\\ & \times
         \Big[
          1 + \sum_{c \in Z^{\Polyn{U}_l} \backslash \Set{b}} e^{-(c-b, y)}
         \Big]^{\gamma}
         \Big[
          {a}_{b'}
           + \sum_{c' \in Z^{\Polyn{F}_l} \backslash \Set{b'}} 
             {a}_{c'} e^{-(c'-b', y)}
         \Big]^{\beta} ,
\end{split}
\end{equation}
where
\begin{eqnarray}
\Delta_{b b'} := \Delta_b^{\Polyn{U}_l} \cap \Delta_{b'}^{\Polyn{F}_l} .
\label{geo:delta}
\end{eqnarray}
Our first and second problems are solved when we find
how to construct
\(\Delta_{b b'}
     \;(\forall b \in Z^{\Polyn{U}_l},
    \ \;\forall b' \in Z^{\Polyn{F}_l})\).

These problems are interpreted in the words of convex geometry.
We introduce several terms and notations 
according to the appendix of Ref.~\cite{math:Oda-1987}.
The \textit{convex polyhedral cone}
for a finite set 
\(S\) in \(n\)-dimensional Euclidean space  is defined by:
\begin{eqnarray}
  C(S) := \SetIn{\sum_{v \in S} r_v v \in \SetReal^n}{
          r_v \in \SetReal_{\geq 0}, \; \forall v \in S} .
\label{geo:cs}
\end{eqnarray}
It is noted that a line is a convex polyhedral cone for
$S = \Set{v, -v}$ with a fixed vector $v$.
Similarly, it is easy to see that an $m$-dimensional linear subspace 
$\SetReal^m\;(m \leq n)$
is also a convex polyhedral cone.

The \textit{dual cone} of a convex polyhedral cone \(C\) is
defined by :
\begin{eqnarray}
\label{geo:dualc}
\DualCone{C} &:=& \SetIn{y \in \SetReal^n}{
          (v, y) \geq 0, \; \forall v \in C} .
\end{eqnarray}
For a finite set \(S\), \(\DualCone{C(S)}\) is also a
convex polyhedral cone.

Let \(Z_b^{P}\) be the set of points translated from \(Z^{P}\)
such that point \(b\) is moved to the origin:
\begin{eqnarray}
Z_b^{P} &:=& \SetIn{c - b \in \SetInteger^{N-1}}{c \in Z^P} .
\end{eqnarray}
Comparing with Eqs.~(\ref{geo:dom}) and (\ref{geo:dualc}), 
we have
\begin{eqnarray}
\Delta_b^P &:=& \DualCone{C(Z_b^{P})} \cap \SetReal_{\geq 0}^{N-1} ,
\end{eqnarray}
and
\begin{eqnarray}
\Delta_{b b'} &=&
  \DualCone{C(Z_b^{\Polyn{U}_l})} \cap \DualCone{C(Z_{b'}^{\Polyn{F}_l})}
  \cap \SetReal_{\geq 0}^{N-1} .
\label{dbb}
\end{eqnarray}

Thus what we have to solve is to find how to construct cones
($C(Z_b^{\Polyn{U}_l})$ and $C(Z_{b'}^{\Polyn{F}_l})$), dual cones
($\DualCone{C(Z_b^{\Polyn{U}_l})}$ and
$\DualCone{C(Z_{b'}^{\Polyn{F}_l})}$),
the intersections of dual cones ($\Delta_{bb'}$) and to see
whether their volume is zero or not.
The algorithms to solve these problems are developed
in combinatorial geometry, which will be discussed
in the next section.

\begin{figure}[!ht]
    \begin{center}
    \includegraphics[width=12cm,clip=true]{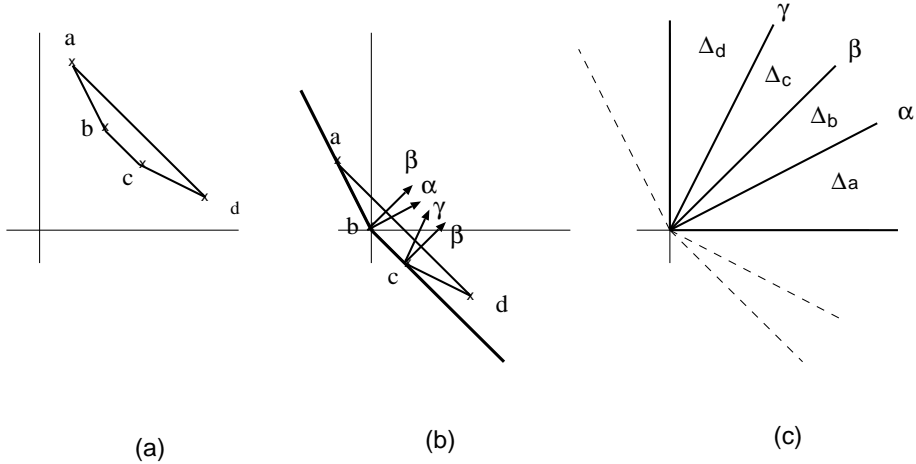}
    \caption{An example of cone and dual cone
             \label{fig:sample} }
    \end{center}
\end{figure}

Fig.~\ref{fig:sample} shows a simple example for
the case of two variables:
\begin{eqnarray}
P(t) &=& t_1 t_2^5 + t_1^2 t_2^3 + t_1^3 t_2^2 + t_1^5 t_2 .
\end{eqnarray}
Fig.~\ref{fig:sample}(a) and (b) show \(Z^P, \; Z_b^{P}\),
respectively.
Vectors \(\alpha\), \(\beta\) and \(\gamma\) satisfies the condition
\((\alpha, v-b) \geq 0\), \((\beta, v-b) \geq 0\),
\((\beta, v-c) \geq 0\) and \((\gamma, v-c) \geq 0\) for all
\(v \in Z^P\).
Fig.~\ref{fig:sample}(c) shows the resultant sub-domains.

Since these cones are in general still too complex
for the integration,
we divide them into a set of simpler objects.
In the case of polytope (bounded polyhedron), 
it can be decomposed as a union
of simplicies, which is called \textsl{triangulation} .
The algorithms of triangulation
is also studied in combinatorial geometry.
Similarly, a cone can be decomposed as a union of
simplicial cones as discussed in the next section.

An \((N-1)\)-dimensional cone \(C(V)\) is 
\textit{simplicial} when
\(V = \Set{v_1, \cdots, v_{N-1}}\)
is a set of
$(N-1)$ linear independent vectors.
Let \(S_{b b'}\) be the set of $(N-1)$ vectors characterizing
simplicial cones such that
\begin{eqnarray}
\Delta_{b b'} = \bigcup_{V \in S_{b b'}} C(V)
\end{eqnarray}
represents a triangulation of $\Delta_{b b'}$.
With this triangulation, Eq.~(\ref{geo:gly}) is expressed by:
\begin{equation}
\begin{split}
G_l =&
      \sum_{b \in Z^{\Polyn{U}_l}}
      \sum_{b' \in Z^{\Polyn{F}_l}}
      \sum_{V \in S_{b b'}}
      \int_0^\infty d^{N-1} y \;
       \theta(y \in C(V)) \;
         e^{-(\nu' + \gamma b +
               \beta b', y)}  \,
\\ & \times
         \Big[
          1 + \sum_{c \in Z^{\Polyn{U}_l} \backslash \Set{b}} e^{-(c-b, y)}
         \Big]^{\gamma}
         \Big[
          {a}_{b'}
           + \sum_{c \in Z^{\Polyn{F}_l} \backslash \Set{b'}} {a}_{c} e^{-(c-b', y)}
         \Big]^{\beta} .
\end{split}
\end{equation}

A point \(y = (y_i) \in C(V)\) is parameterized by
\textsl{barycentric coordinate}:
\begin{eqnarray}
y_i &=& \sum_{j=1}^{N-1} (v_j)_i u_j ,
\qquad ( u_j \in \SetReal_{\geq 0}, \quad j=1,\cdots, N-1) .
\end{eqnarray}
Since \(\Set{v_j}\) is a set of \((N-1)\) linear independent vectors,
the correspondence between
\(\Set{y_i}\) and \(\Set{u_j}\) is one-to-one.
We regard \(V = (v_1, \cdots, v_{N-1})\) 
being a \((N-1)\times(N-1)\)
matrix made arranging column vectors \(\Set{v_j}\).
Point \(y\) is expressed by \(y = V u\) 
in this matrix notation.
We subsequently change the variable from \(y\) to \(u\)
then from \(u\) to \(z\), where \(z_j = e^{-u_j}\).
Jacobian is \(| \det V|/\prod_j z_j\).
We finally obtain:
\begin{equation}
\label{geo:glz}
\begin{split}
G_l =&
      \sum_{b \in Z^{\Polyn{U}_l}}
      \sum_{b' \in Z^{\Polyn{F}_l}}
      \sum_{V \in S_{b b'}}
      | \det V |
      \int_0^1 d^{N-1} z \;
         \prod_{j=1}^{N-1}
         z_j^{(\nu' + \gamma b +
               \beta b', v_j)-1}  \,
\\ & \times
         \Big[
          1 + \sum_{c \in Z^{\Polyn{U}_l} \backslash \Set{b}}
               \prod_{j=1}^{N-1} z_j^{(c-b, v_j)}
         \Big]^{\gamma} \,
         \Big[
          {a}_{b'}
           + \sum_{c \in Z^{\Polyn{F}_l} \backslash \Set{b'}} {a}_{c}
               \prod_{j=1}^{N-1} z_j^{(c-b', v_j)}
         \Big]^{\beta} .
\end{split}
\end{equation}
The original variables are expressed by:
\begin{eqnarray}
& &
t_j \Eq e^{-y_j}
\Eq e^{-(V u)_j}
\Eq \prod_k z_k^{(v_k)_j} ,
\end{eqnarray}
with Jacobian
\(|\det V| {\prod_j t_j} / {\prod_k z_k}\) .

As it will be shown in the next section,
we can take vectors \(\Set{v_j}\) on the integer lattice
\(v_j \in \SetInteger_{\geq 0}^{N-1} \backslash \{0\}\).
Since $(c-b,y) \geq 0$ for all 
$y \in \Delta_b^P$ and $(c-b)\in \SetInteger^{N-1}$,
\((c-b, v_j)\) is non-negative integer.
Thus the sub-expressions in the brackets
of Eq.~(\ref{geo:glz})
are polynomials of \(z\).

For fixed \(k\),
let us consider the scale transformation of vector \(v_k\)
and variable \(z_k\) defined
by \(v_k \rightarrow \lambda_k v_k\) and
\(z_k \rightarrow z_k^{1/\lambda_k}\) for \(\lambda_k > 0\).
It is easy to see that integration \(G_l\) is invariant
under this transformation.
We select \(v_k\) on the integer lattice as nearest possible to the
origin.

Now all the problems are converted to the following ones
in convex and combinatorial geometry:
\begin{enumerate}
\item to construct intersections of cones
      \(\Delta_{b b'} =
      \DualCone{C(Z_b^{\Polyn{U}_l})} \cap \DualCone{C(Z_{b'}^{\Polyn{F}_l})}
      \cap \SetReal_{\geq 0}^{N-1}\) 
      for all \(b \in Z^{\Polyn{U}_l}\) and
      \(b' \in Z^{\Polyn{F}_l}\).
\item to triangulate \(\Delta_{b b'}\) into 
      \(\cup_{V \in S_{b b'}} C(V)\).
\end{enumerate}

\section{Algorithms}

\subsection{Construction of cones}

First we rewrite \(\Delta_{bb'}\) into a different form.
The following relations hold 
for convex polyhedral cones
\(C,\; C_1\) and \(C_2\), and for sets of finite points \(S_1\) and \(S_2\)
(see. Ref.~\cite{math:Oda-1987} Appendix):
\begin{eqnarray}
C(S_1 \cup S_2) &=& C(S_1) + C(S_2) ,
\\
\DualDualCone{C} &=& C ,
\\
\DualCone{C_1} \cap \DualCone{C_2} &=& \DualCone{(C_1+C_2)} ,
\end{eqnarray}
where addition `\(+\)' represents \textit{Minkowsky sum}
defined by:
\begin{eqnarray}
C_1 + C_2 \;:=\; \SetIn{a + b}{a \in C_1, \; b \in C_2} .
\label{minkowsky}
\end{eqnarray}
Let \(E^m\) be the standard basis of Euclidean space \(\SetReal^m\),
that is 
\begin{eqnarray}
&& 
E^m := \Set{e_1, e_2, \cdots, e_m} ,
\\ &&
   e_1 = (1,0,0,\cdots,0), \;
   e_2 = (0,1,0,\cdots,0), \; \cdots  .
\end{eqnarray}
Set \(\SetReal_{\geq 0}^{m}\) is represented by:
\begin{eqnarray}
\SetReal_{\geq 0}^{m} &=& \DualCone{(\SetReal_{\geq 0}^m)}
   \Eq C(E^{m}) \Eq \DualCone{C(E^{m})} .
\end{eqnarray}
With these relations, Eq.~(\ref{dbb}) is rewritten to:
\begin{equation}
\begin{split}
\Delta_{bb'} &=
  \DualCone{C(Z_b^{\Polyn{U}_l})} \cap \DualCone{C(Z_{b'}^{\Polyn{F}_l})}
  \cap \DualCone{C(E^{N-1})}
\\ &=
  \DualCone{\{ C(Z_b^{\Polyn{U}_l})+ C(Z_{b'}^{\Polyn{F}_l}) + C(E^{N-1})\}}
\\ &=
  \DualCone{C(Z_b^{\Polyn{U}_l} \cup Z_{b'}^{\Polyn{F}_l} \cup E^{N-1} )}
   .
\end{split}
\end{equation}
Thus we obtain
\begin{eqnarray}
\Delta_{bb'} &=& \DualCone{C(Z_{bb'})}
  ,
\end{eqnarray}
where
\begin{eqnarray}
Z_{bb'} &:=& Z_b^{\Polyn{U}_l} \cup Z_{b'}^{\Polyn{F}_l} \cup E^{N-1} .
\end{eqnarray}
What we should know is how to construct a convex polyhedral cone
for a given finite set and how to construct its dual cone.

Let us introduce some terms used in convex geometry.
A \textit{convex hull} of a subset \(T\) of \(\SetReal^m\) 
is the smallest convex set including \(T\) and is
denoted by \(\ConvexHull(T)\).
When \(T\) is a finite set of half-lines starting from the origin,
\(\ConvexHull(T)\) is a convex polyhedral cone.
When \(T\) is a finite set of points, 
\(\ConvexHull(T)\) is a polytope.
When \(T\) is equal to the set \(Z^P\) corresponding to a polynomial \(P\),
\(\ConvexHull(Z^P)\) is called \textit{Newton polytope}
of \(P\), which plays important role for analyzing
properties of a multi-variate polynomial.

Let \(K\) be a \(k\)-dimensional polytope or convex polyhedral cone
in \(m\)-dimensional Euclidean space.
A hyperplane \(H\) is called to \textit{support} a set \(K\) 
at a point \(x \in K\)
if \(x \in H\) and \(K\) is included in one of the two closed 
half-spaces limited by \(H\).
That is, when a vector \(u\) is normal to \(H\),
either \( (u, y-x) \geq 0 (\forall y \in K)\) or
       \( (u, y-x) \leq 0 (\forall y \in K)\) is true.
A subset \(F\) of \(K\) is called a \textit{faces} of \(K\)
if either \(F = \emptyset\) or \(F = K\), or if
there exists a supporting hyperplane \(H\) of \(K\) such that
\(F = H \cap K\).
A face of convex polyhedral cone or polytope is also 
a convex polyhedral cone or a polytope, respectively.
A face is called a \textit{vertex}, an \textit{edge}
or a \textit{facet} when it is 0-, 1- or \((k-1)\)-dimensional, 
respectively.
The set of \(j\)-dimensional faces is denoted by \(\Face_j(K)\)
and the set of all faces by
\(\Face(K) := \Set{\emptyset} \cup_{j=0}^{k} \Face_j(K)\).
In the case of a convex polyhedral cone, \(\Face_0\) is \(\Set{0}\)
if \(\Face_0 \neq \emptyset\).
It is noted that some special convex polyhedral cones,
such as lines or the whole space, 
have no vertices.
A face of a polytope is expressed by a set of vertices
included in the face.
On the other hand, an edge of convex polyhedral cone
is expressed by a non-zero vector laying on the edge.
A higher than one-dimensional face 
is expressed by a set of edges
included in the face.
Let \(f\) and \(g\) be two faces of \(K\).
When \(f\) is a face of \(g\), this relation is denoted by \(f \prec g\).
The set \(\Face(K)\) forms an \textit{abstract complex} structure 
in terms of this binary relation \(\prec\)
(see. Ref.~\cite{math:Oda-1987} Proposition A.5 and A.16).

The convex hull problem of a polytope is to find \(\Face(\ConvexHull(S))\) 
for a set of finite points \(S\).
This problem is popular and well studied
in computational geometry
(see, for example, Ref.~\cite{math:Edelsbrunner-1987}).
Several software packages have also been developed
to solve these problems~\cite{comp:QHull,comp:Hull}.

Our problem is to construct an algorithm to find
\(\Face(C(S))\) for a convex polyhedral cone $C(S)$.
There will be two approaches.
One is to modify directly an algorithm
for a polytope to one for a convex polyhedral cone.
It is not so difficult and
we have prepared one for our test program
based on the algorithm described in 
Ref.~\cite{math:Edelsbrunner-1987}.
The other is to use the output of algorithms for polytopes.
We discuss here the latter approach.

Let us consider a correspondence between a convex polyhedral cone
\(C(S)\) and a polytope \(\ConvexHull(S)\)
defined for the same set of finite points
\(S \subset \SetReal^m\).
The polytope $\ConvexHull(S)$ is expressed by:
\begin{eqnarray}
\ConvexHull(S) = \SetIn{ \sum_{u \in S} \lambda_u u }{
                    \lambda_u \geq 0, \sum_{u \in S} \lambda_u = 1 } .
\label{alg:convs}
\end{eqnarray}
Since our set of points \(Z_{bb'}\) includes the origin \(0\),
we assume that \(0 \in S\).
Let \(\Face(\ConvexHull(S))\)
be the set of faces obtained by a convex hull algorithm for a polytope.
Comparing (\ref{geo:cs}) with (\ref{alg:convs}), 
\(C(S)\) is expressed by:
\begin{equation}
\begin{split}
C(S) &=  \SetIn{\lambda x}{
                x \in \ConvexHull(S),\; 
                \lambda \in \SetReal_{\geq 0}}
\\   &=  \SetIn{\sum_{v \in \Face_0(\ConvexHull(S))} \lambda_v v}{
                \lambda_v \in \SetReal_{\geq 0}, \; 
               \forall v \in \Face_0(\ConvexHull(S))}.
\end{split}
\end{equation}
Although \(S\) includes the origin,
it is possible that the origin is not a vertex of \(\ConvexHull(S)\).
In this case, 
there exists a line segment \(L\) 
passing through the origin
and being included in both \(\ConvexHull(S)\) and \(C(S)\).
Let the line segment \(L\) be expressed 
by \(L = \SetIn{\lambda v}{a \leq \lambda \leq b}\) 
with some non-zero vector \(v\) and real numbers \(a < 0\) and \(b > 0\).
For any \(y \in \DualCone{C(S)}\), 
condition \(\lambda (v, y) \geq 0\) is satisfied 
for both positive and negative value of 
\(\lambda\), and then \((v, y) = 0\).
So \(\DualCone{C(S)}\) is included in a hyperplane
perpendicular to vector \(v\).
It means that the dimension of \(\DualCone{C(S)}\) is less than \(m\),
and this sub-domain does not contribute to the integration in our problem.

Now we assume the origin is a vertex of \(\ConvexHull(S)\).
Then we can assume the origin is also a vertex of \(C(S)\).
If not so, any hyperplane
passing through the origin does not support \(C(S)\) nor \(\ConvexHull(S)\),
which is a contradiction to the first assumption.
These convex polyhedral cones are called \textit{strongly convex}.
Let \(\Face_j(C(S))\) be the set of \(j\)-dimensional faces of \(C(S)\)
and \(\Face_j'(\ConvexHull(S))\) be the set of \(j\)-dimensional faces of 
\(\ConvexHull(S)\) including the origin : 
\(\Face_j'(\ConvexHull(S)) = \SetIn{f \in \Face_j(\ConvexHull(S))}{0 \in f}\).
Let a face \(f \in \Face_j'(\ConvexHull(S))\) be defined by 
\(f = H \cap \ConvexHull(S)\) with
a supporting half-plane \(H\).
As \(H\) also supports \(C(S)\) at the origin,
a subset \(g = H \cap C(S)\) of $C(S)$
is a \(j\)-dimensional face of \(C(S)\).
It is easy to show that
this correspondence between a face in
\(\Face_j'(\ConvexHull(S))\) and one in \(\Face_j(C(S))\)
is one-to-one and preserves the relation \(\prec\).
It implies that the structure of abstract complex
of \(\Face(C(S))\) is the same as one of 
\(\Face'(\ConvexHull(S))\).
As the result, we obtain the following
algorithm of constructing \(C(S)\) in \(\SetReal^m\):
\begin{enumerate}
\item Construct \(\Face(\ConvexHull(S))\) from \(S\)
      by a convex hull algorithm for a polytope.
\item If \(0 \not\in \Face_0(\ConvexHull(S))\) then we ignore
      \(\DualCone{C(S)}\), since it does not contribute to the
      integration.
\item Construct a subset of faces \(\Face'(\ConvexHull(S))\) selecting 
      ones including the origin.
\item Interpret \(\Face'(\ConvexHull(S))\) 
      as the set of faces \(\Face(C(S))\).
\end{enumerate}

The dual cone of \(C(S)\) is constructed
from \(\Face(C(S))\) by the dual operation
shown by Proposition A.6 of Ref.~\cite{math:Oda-1987}.
In our case, it is sufficient to know the set of edges
\(\Face_1(\DualCone{C(S)})\), since \(\DualCone{C(S)}\) is expressed by:
\begin{eqnarray}
\DualCone{C(S)}
     &=& \SetIn{\sum_{u \in \Face_1(\DualCone{C(S)})} \mu_u u}{
                \mu_u \in \SetReal_{\geq 0}, \; 
               \forall u \in \Face_1(\DualCone{C(S)})},
\label{dcsv}
\end{eqnarray}
where \(u \in \Face_1(\DualCone{C(S)})\) means to select a
non-zero vector \(u\) for each edge such that \(u\) lies
on the edge.
If \(\dim C(S) = m\),
an edge of \(\DualCone{C(S)}\) is obtained as 
a half-line starting from the origin and parallel to
the normal vector of a facet of \(C(S)\).
If \(\dim C(S) < m\), \(\DualCone{C(S)}\) is not strongly convex
and the construction of \(\Face_1(\DualCone{C(S)})\) 
is slightly complicated.
However, we do not need to consider this case,
as discussed before.

We show that when \(S\) is a set of integer vectors,
a vector \(u \in \Face_1(\DualCone{C(S)})\)
can be selected as an integer vector.
Since vertices of \(\ConvexHull(S)\) are included in the original
set of points \(S\), 
we can select integer vectors to represent the edges of \(C(S)\).
A facet of \(C(S)\) is a subset of hyperplane spanned by some
of these vectors.
A vector representing an edge of \(\DualCone{C(S)}\) is a vector
normal to the facet.
The normal vector is obtained by solving a system of homogeneous 
linear equations.
The vector \(u\) is solved as
a vector \(w\) multiplied by the inverse matrix
of an integer matrix \(M\).
The inverse matrix is constructed by the minor of \(M\)
divided by the determinant \(\det M\),
where the elements of the minor are integers.
By adjusting the normalization of vector \(w\), 
we can cancel the denominator \(\det M\).
In this way, we can take \(v\) as an integer vector.

As the result, \(\Delta_{bb'}\) can be expressed with a set of
integer vectors 
\(\Set{v_1, \cdots, v_m} \subset \SetInteger_{\geq 0}^{N-1}\) by:
\begin{eqnarray}
\Delta_{bb'} &=& \SetIn{\sum_{i} r_i v_i}{r_i \geq 0} .
\end{eqnarray}

With the fact that
our set of points \(Z_{bb'}\) includes \(E^{N-1}\) as a subset,
we can accelerate the algorithm in the following cases:
\begin{itemize}
\item Case where there exists a non-zero vector
      \(v \in Z_b^{\Polyn{U}_l} \cup Z_{b'}^{\Polyn{F}_l}\)
      and \(v \in \SetReal_{\leq 0}^{N-1}\).
      In this case, negated vector \(-v\) is expressed 
      by a linear combination
      of vectors in \(E^{N-1}\) with non-negative coefficients.
      It means that 
      \(C(Z_b^{\Polyn{U}_l} \cup Z_{b'}^{\Polyn{F}_l} \cup E^{N-1})\)
      includes line segment connecting \(v\) and \(-v\) and thus
      the origin is not the vertex.
      So we can discard such a configuration.

\item Case where there exists a non-zero vector
      \(v \in Z_b^{\Polyn{U}_l} \cup Z_{b'}^{\Polyn{F}_l}\)
      and \(v \in \SetReal_{> 0}^{N-1}\).
      In this case, vector \(v\) is an interior point of
      \(C(Z_b^{\Polyn{U}_l} \cup Z_{b'}^{\Polyn{F}_l} \cup E^{N-1})\)
      and is not on the edge of the cone.
      So we can eliminate \(v\) from the input set of points.
      That is, the input set of points
      \(Z_b^{\Polyn{U}_l} \cup Z_{b'}^{\Polyn{F}_l} \cup E^{N-1}\)
      can be replaced by
      \(\{(Z_b^{\Polyn{U}_l} \cup Z_{b'}^{\Polyn{F}_l} ) \backslash
          \SetReal_{\geq 0}^{N-1} \} \cup \Set{0}
          \cup E^{N-1}\).
      This replacement decreases the number of input points,
      which accelerate the convex hull algorithm.
\end{itemize}
These conditions are easily checked before applying the convex
hull algorithm.

\subsection{Triangulation of convex polyhedral cones}

We discuss here how to triangulate convex polyhedral cones.
Since a convex polyhedral cone is not bounded,
we convert it to a bounded polytope in order to make the problem
easier to handle.
Let \(H\) be a hyperplane perpendicular
to vector \(\VOne := (1, 1, \cdots, 1)\)
and passing through a point 
\(w \VOne\) for \(w \in \SetReal_{> 0}\).
As \(\Delta_{b b'} \subset \SetReal_{\geq 0}^{N-1}\),
this cone can be truncated to a polytope 
in cutting by \(H\).
Then the resulting polytope \(\Delta_{b b'}'\)
is expressed by:
\begin{eqnarray}
\Delta_{b b'}' &=& \SetIn{\sum_j r_j v_j'}{
                  0 \leq r_j \leq 1, \; \sum_j r_j \leq 1} ,
\\
v_j' &:=& v_j w/(\VOne, v_j) ,
\qquad
(\VOne, v_j') = w .
\end{eqnarray}
Since \(v_j\) is an integer vector, \((\VOne, v_j)\) is integer.
By selecting \(w\) as the least common multiplier 
\(\LCM\SetIn{(\VOne, v_j)}{\forall j}\),
we can take all of \(\Set{v_j'}\) as integer vectors.

The triangulation of \((N-1)\)-dimensional polytope is not evident.
We restrict ourselves to triangulation methods such that
no new vertices are added.
Even with this restriction, the number of resulting simplicies
is not unique.
\begin{figure}[!ht]
    \begin{center}
    \includegraphics[width=12cm,clip=true]{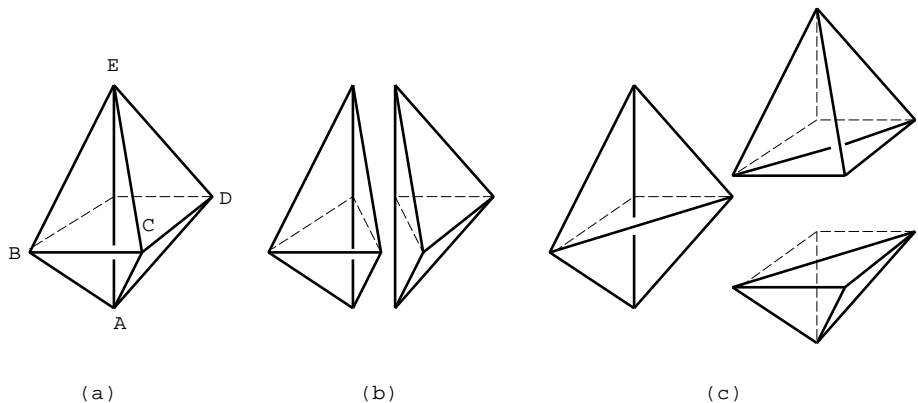}
    \caption{An example of different triangulations
             \label{fig:tri3} }
    \end{center}
\end{figure}
The following example (Fig.~\ref{fig:tri3}) shows
the case where the different numbers of simplicies
are produced by different triangulations.
Let \(A, B, C, D, E\) be points in the 3-dimensional space
defined by:
\begin{eqnarray}
\begin{array}{l}
A \Eq (0, 0, 0) , \quad
B \Eq (1, 0, 1) , \quad
C \Eq (1, 1, 1) , \quad
\\
D \Eq (0, 1, 1) , \quad
E \Eq (0, 0, 2) .
\end{array}
\end{eqnarray}
The convex hull of these points is shown by Fig.~\ref{fig:tri3}(a).
In cutting the convex hull
by a triangle \((A, C, E)\), we obtain two simplicies
(Fig.~\ref{fig:tri3}(b)).
On the other hand, there are three simplicies
when it is cut
by three triangles
\((A, B, D)\), \((E, B, D)\) and \((B,C,D)\)
(Fig.~\ref{fig:tri3}(c)).
We do not know triangulation algorithms which produce
the minimum number of simplicies.
We use the following algorithm.
\begin{enumerate}
\item If the number of vertices of the input
      \(d\)-dimensional polytope \(P\) is \(d+1\) then
      \(P\) is a simplex.
      The algorithm returns the set \(\Set{P}\).
      This condition is always satisfied when \(d=1\)
      (\(P\) is a line segment).
\item Pick up one vertex \(u\) from \(P\).
      Remove \(u\) and all interior points of all line
      segments included in $P$ and
      starting from $u$.
      The resulting geometric object is the union of
      polytopes \(Q_j\; (j=1,\cdots,m)\) of dimension \(d-1\),
      which are $(d-1)$-dimensional faces not including $u$.
\item For all \(j = 1, \cdots, m\),
      apply triangulation algorithm to \(Q_j\) recursively,
      and set \(S_j\) of simplicies is obtained.
\item Make the union \(S = \cup_j S_j\).
\item Construct a \(d\)-dimensional simplicies by adding
      \(u\) to each (\(d-1\))-dimensional simplex in \(S\).
      The algorithm returns the set of all constructed simplicies.
\end{enumerate}
It is noted that this algorithm never falls into the infinite loop.
Since step 2 removes at least one face, the number of recusive
applications of the algorithm is bounded by the number of faces
of the polytope.
The number of resulting simplicies depends on how vertex \(u\) is
selected in step~2.
One choice is to select \(u\) when it attaches the
maximum number of edges.
This condition produces relatively smaller number of resulting
simplicies, which result will be shown in the next section.

For the triangulation of a cone, 
one must select the origin
for \(u\) in the step~2 at the first time before calling subsequent
recursive procedure.
This selection guarantees that each simplex includes the origin
as one of its vertices.
Let \(s_v\) be one of the obtained simplicies
and characterized by
a set of \(N-1\) linearly independent vectors \(\Set{v_j}\) as:
\begin{eqnarray}
s_v &=& \SetIn{ \sum_j r_j v_j}{
                \sum_j r_j \leq 1, \; 0 \leq r_j \; (\forall j)} .
\end{eqnarray}
From this simplex, a simplicial cone \(s_v'\) is obtained
by extending the range
of the values of \(\Set{r_j}\):
\begin{eqnarray}
s_v' &=& \SetIn{ \sum_j r_j v_j}{ 0 \leq r_j \; (\forall j)} .
\end{eqnarray}
It is noted that vector \(v_j\) is an integer vector,
for it is one of the
vertices in the polytope before triangulation.

\section{A test implementation}

We have prepared a program in order to confirm the correctness
of our method.
For the convex hull method we adopted incremental algorithm
described in Ref.~\cite{math:Edelsbrunner-1987} and
modified it for convex polyhedral cones.
In order to avoid the accuracy problem of numerical data,
all numerical operation are performed in integer arithmetic.
We adopted program language \texttt{python} which is equipped with
arbitrary long integer arithmetic.
Input data is prepared by the program package described in 
Ref.~\cite{phys:Ueda-Fujimoto-2008}.
The output integrand of our program is passed to the same program package,
which continues the subsequent divergence separation
and integration steps.

The procedure of constructing convex hull is compared with
\texttt{qhull} package \cite{comp:QHull}.
The triangulation procedure is checked by computing the volume
of resulting simplicies.
Integrated values are compared with ones in published articles
in several cases.
We show the number of decomposed sectors in Table~\ref{ex:tab}.
Column ``H'' is cited from Ref.~\cite{phys:Heinrich-2008} and
columns ``A'', ``B'', ``C'', ``S'' and ``X'' from 
Ref.~\cite{phys:Smirnov-Tentyukov-2008}
\begin{table}[!ht]
\caption{The number of sectors}
\label{ex:tab}
\begin{center}
{\small
\begin{tabular}{|l|ccccc|c|c|}
 \hline
Diagram
    &    A &      B &      C &      S &      X & H & This \\
    &      &        &        &        &        &   & method \\
 \hline
Bubble
    &   2  &      2 &      2 &    --- &      2 &    --- &      2 \\
Triangle
    &   3  &      3 &      3 &    --- &      3 &    --- &      3 \\
Box
    &   12 &     12 &     12 &     12 &     12 &    --- &     12 \\
\hline
Tbubble
    &   58 &     48 &     48 &    --- &     48 &    --- &     48 \\
Double box, \(p_i^2 = 0\)
    &  775 &    586 &    586 &    362 &    293 &    282 &    266 \\
Double box, \(p_4^2 \neq 0\)
    & ---  &    --- &    --- &    --- &    --- &    197 &    186 \\
Double box, \(p_i^2 = 0\)
    & 1138 &    698 &    698 &    --- &    395 &    --- &    360 \\
~~ nonplanar  & &  & & & & & \\
D420
    & 8898 &    564 &    564 &    180 &      F &    --- &    168 \\
\hline
3 loop vertex (A8)
    & ---  &    --- &    --- &    --- &    --- &    684 &    684 \\
Triple box
    &    M & 114256 & 114256 &  22657 &  10155 &    --- &   6568 \\
 \hline
 \end{tabular}
}
\end{center}
\end{table}

This test program is too slow for the practical purposes.
We are planning to prepare a practical implementation in \texttt{C++}.

\section{Conclusion}

We proposed a new method of IR factorization 
in sector decomposition
employing a geometric interpretation of the problem.
The original problem is converted into a set of problems in
convex geometry: (1) construction of intersection among dual
cones of convex polyhedral cones corresponding to each
polynomial in the original problem and (2) triangulation of
them.
They are solved by the algorithms in combinatorial geometry.
This is a deterministic method and never falls into a infinite
loop. 
The number of resulting sectors depends on the algorithm of
triangulation. 
Our test implementation of this method shows the smaller number
of sectors compared to other methods based on iterated
decomposition.

\vspace{1em}
\noindent
\textbf{Acknowledgments}

The authors wish to express their thanks to the members
of Minami-tateya group for the useful discussions.
This work is supported in part by
Ministry of Education, Science, and Culture, Japan
under Grant-in-Aid No.20340063 and No.21540286.


\end{document}